\documentclass{article}

\begin{document}

\newcommand{\q}{\ensuremath{q}}
\newcommand{\qs}{\ensuremath{q^2}}
\newcommand{\so}{\ensuremath{SO_{q}(3)}}                
\newcommand{\invR}{\ensuremath{\frac{1}{R}}}    
\newcommand{\invRn}[1]{\ensuremath{\frac{1}{R^{#1}}}}   
\newcommand{\dl}{\ensuremath{\partial}}
\newcommand{\dbar}{\ensuremath{\overline{\partial}}}
\newcommand{\kron}[2]{\ensuremath{\delta^{#1}_{#2}}}
\newcommand{\met}[2]{\ensuremath{\gamma^{#1#2}}}
\newcommand{\imet}[2]{\ensuremath{\gamma_{#1#2}}}
\newcommand{\eps}[3]{\ensuremath{\epsilon_{#1}\,^{#2#3}}}
\newcommand{\ieps}[3]{\ensuremath{\epsilon_{#1#2}\,^{#3}}}
\newcommand{\sq}{\ensuremath{\sqrt{q}}}        
\newcommand{\isq}{\ensuremath{\frac{1}{\sq}}}  
\newcommand{\iq}{\ensuremath{\frac{1}{q}}}  
\newcommand{\R}[4]{\ensuremath{\tilde{R}^{#1#2}_{#3#4}}}
\newcommand{\Rinv}[4]{\ensuremath{\tilde{R}^{-1}\,^{#1#2}_{#3#4}}}
\newcommand{\axre}{\ensuremath{\hat{\mathcal{A}}^{X}_{R^{3}_{q}}}}
\newcommand{\dxre}{\ensuremath{\hat{\mathcal{D}}^{X}_{R^{3}_{q}}}}
\newcommand{\axr}{\ensuremath{\mathcal{A}^{X}_{R^{3}_{q}}}}
\newcommand{\dxr}{\ensuremath{\mathcal{D}^{X}_{R^{3}_{q}}}}
\newcommand{\invmu}{\ensuremath{\mu^{-1}}}
\newcommand{\X}[1]{\ensuremath{X_{#1}}}
\newcommand{\qn}[1]{\ensuremath{q^{#1}}}
\newcommand{\yq}[2]{\ensuremath{Y_{q}\,^{#1}_{#2}}}
\newcommand{\QP}{\ensuremath{K^{2}P^{2}}}
\newcommand{\RY}[3]{\ensuremath{R^{#1}\yq{#2}{#3}}}
\newcommand{\nq}[1]{\ensuremath{[#1]_{q}}}
\newcommand{\cosq}{\ensuremath{\cos_{q}}}
\newcommand{\sinq}{\ensuremath{\sin_{q}}}
\newcommand{\expq}{\ensuremath{\exp_{q}}}
\newcommand{\half}{\ensuremath{\frac{1}{2}}}
\newcommand{\suminf}[1]{\ensuremath{\sum_{#1=0}^{\infty}}}
\newcommand{\Rexp}[2]{\ensuremath{R^{\,l+#1}\,\expq(-q^{#2}\beta R)
\,\yq{l}{m}}}
\newcommand{\ga}{\ensuremath{\gamma}}
\newcommand{\poch}[3]{\ensuremath{(#1;#2)_{#3}}}

\bigskip \medskip \begin{center} \large {\bf A \q-deformation
of the Coulomb Problem} \\

 \bigskip \bigskip \normalsize

James Feigenbaum\footnote{Supported by an NSF fellowship; e-mail
jafeigen@yukawa.uchicago.edu}
and Peter G. O. Freund\footnote{Supported in part by NSF grant PHY-91-23780;
e-mail freund@yukawa.uchicago.edu}\\ {\it Enrico Fermi Institute,
Department of Physics \\
University of Chicago, Chicago, IL  60637} \\

 \end{center}
 \bigskip
 \centerline{ABSTRACT}
 \begin{quote}

The algebra of observables of $SO_{q}(3)$-symmetric quantum mechanics is
extended to include the inverse $\frac{1}{R}$ of the radial coordinate
and used to obtain eigenvalues and eigenfunctions of a \q-deformed Coulomb
Hamiltonian.

 \end{quote}

 \newpage

\section{Introduction}

Much work has been done recently to explore the
\so -symmetric quantum mechanics developed in \cite{WZ}
and \cite{OZ}.  In particular,
a lot is known about the \q-deformations of
the harmonic oscillator.  The other nontrivial soluble
problem in ordinary quantum mechanics is the Coulomb
problem, but for that one needs some notion of an inverse
radius.  Weich \cite{Weich} considered a \q-deformed Coulomb
potential,
defining \invR\ in a manner dependent
upon a particular Hilbert space representation.
This differs from the more standard "wave-function" type
approaches used in investigations of the oscillator
(Refs. \cite{Fiore} - \cite{C-W2} for example).

Here we approach this problem by defining \invR\ as an actual
element of the algebra of observables, thereby achieving
representation-independence.
Since $X^{2} = R^{2}$ is
already defined, we can then also define $R$ as well as all its
integral powers.  A study of the action of
momentum operators on powers of R then helps to bring
out the interpretation of these operators as symmetric
\q-derivatives.

Using this definition of \invR , a self-adjoint symmetric
\q-deformation of the Coulomb Hamiltonian can be found which
shares the $n^{2}$-fold degeneracy of the undeformed Hamiltonian
for the \q-analog of bound states.  As in
\cite{Weich}, we obtain a Balmer-type spectrum for these
states with
\begin{equation}
        E_{n} = - \left( \frac{\alpha}{\nq{n}} \right) ^{2},
        \label{eq:Balmer}
\end{equation}
where the symmetric \q-analog of n is defined as
\begin{equation}
        \nq{n} = \frac{ \qn{n} - \qn{-n}}{q - \qn{-1}}.
        \label{eq:qan}
\end{equation}
In addition, we also obtain positive-energy wave functions and
a candidate \q-Coulomb $S$-matrix.

The paper is structured as follows. After a brief review of  \so -symmetric
quantum mechanics (Section 2), we set up the formalism for dealing with
the $q$-Coulomb problem (Sections 3-5), which is then treated in Section 6.
The detailed proofs
of some statements made in the text are deferred to five Appendices.

\section{\so -symmetric quantum mechanics}

We build upon the algebra of observables as it is defined
in Refs. \cite{C-W} and \cite{Weich}.
The \q-deformed metric and Levi-Civit\`{a} tensors are
defined as follows:

\begin{displaymath}
\met{i}{j} \equiv \imet{i}{j} \equiv %
\left[ \begin{array}{ccc}
0       &       0       &       \frac{1}{\sqrt{q}}      \\
0       &       1       &       0                               \\
\sqrt{q}        &       0       &       0               \end{array}
\right]
\end{displaymath}

\begin{displaymath}
\eps{1}{i}{j} \equiv \ieps{i}{j}{1} \equiv %
\left[ \begin{array}{ccc}
0       &       \frac{1}{\sq}   &       0               \\
-\sq    &               0               &       0               \\
0       &               0               &       0       \end{array}
\right]
\end{displaymath}

\begin{displaymath}
\eps{2}{i}{j} \equiv \ieps{i}{j}{2} \equiv %
\left[ \begin{array}{ccc}
0       &               0               &       -1              \\
0       &        \isq - \sq             &       0               \\
1       &               0               &       0       \end{array}
\right]
\end{displaymath}

\begin{displaymath}
\eps{3}{i}{j} \equiv \ieps{i}{j}{3} \equiv %
\left[ \begin{array}{ccc}
0       &               0               &       0               \\
0       &               0               &       \isq            \\
0       &               -\sq            &       0       \end{array}
\right]
\end{displaymath}

These give rise to an R-matrix:
\begin{displaymath}
        \R{i}{j}{k}{l} = q\kron{i}{k}\kron{j}{l} -
        \eps{a}{i}{j}\ieps{k}{l}{a}
        + (\iq - 1)\met{i}{j}\imet{k}{l}
\end{displaymath}
which is a solution to the Yang-Baxter equation
\begin{displaymath}
        \R{i}{j}{a}{b}\R{b}{k}{c}{n}\R{a}{c}{l}{m} =
        \R{j}{k}{d}{e}\R{i}{d}{l}{f}\R{f}{e}{m}{n},
\end{displaymath}
and has the inverse
\begin{displaymath}
        \Rinv{i}{j}{k}{l} = \iq\kron{i}{k}\kron{j}{l} -
        \eps{a}{i}{j}\ieps{k}{l}{a}
        + (q - 1)\met{i}{j}\imet{k}{l}.
\end{displaymath}

We use the metric and Levi-Civit\`{a} tensors to define
scalar and vector products as for the undeformed
tensors:  $A \cdot B = \met{i}{j} A_{i}B_{j}$ and
$[A \times B]_{k} = \eps{k}{i}{j} A_{i}B_{j}$.

\axr is the \so -covariant *-algebra defined by the
generators $X_{1}$, $X_{2}$, $X_{3}$
subject to the relations
\begin{displaymath}
        [X\times X]_{k} = 0
\end{displaymath}
and
\begin{displaymath}
        X_{i}^{\ast} = \met{i}{j}X_{j}.
\end{displaymath}
$X^{2} \equiv X \cdot X$ is then real and central in
this algebra.  In the $q = 1$ limit, $X_{1}$ and $X_{3}$
correspond to
$\frac{1}{\sqrt{2}}(X \pm Y)$ while $X_{2}$ is $Z$.
The space of wave functions in harmonic-oscillator
treatments of \so -symmetric quantum mechanics is
an appropriate subspace of \axr.

One also considers an \so -covariant *-algebra \dxr
of operators on \axr, whose generators are
the $X_{i}$, derivative
operators $\dbar_{i}$, and a scaling operator $\mu$.
The $X_{i}$ act on \axr by left multiplication.
$\mu$ is defined such that $\mu (1) = 1$; and for all
$f \in \axr$, $\mu (X_{i}f) = qX_{i}\mu (f)$.
$\dbar_{i}$ is defined such that $\dbar_{i} (1) = 0$;
and for all $f \in \axr$,
\begin{displaymath}
        \dbar_{i} (X_{i}f) = [\imet{i}{j}
                + \iq \R{k}{l}{i}{j}X_{k}\dbar_{l}]f.
\end{displaymath}
The generators of \dxr\ then obey the relations:
\begin{displaymath}
        \mu X_{i} = qX_{i}\mu,
\end{displaymath}

\begin{displaymath}
        \mu \dbar_{i} = \iq \dbar_{i} \mu,
\end{displaymath}

\begin{displaymath}
        \dbar_{i} X_{j} = \imet{i}{j} + \iq \R{k}{l}{i}{j} X_{k} \dbar_{l}
\end{displaymath}

\begin{displaymath}
        [\dbar \times \dbar]_{k} = 0.
\end{displaymath}
One also defines an inverse of $\mu$:
\begin{displaymath}
        \invmu \equiv \mu [1
                + q^{-2}(1 - q^{2})X \cdot \dbar
                + q^{-3}(1 - q)^{2}X^{2}\dbar^{2}].
\end{displaymath}
In addition, there is a conjugate set of
derivative operators in \dxr,
\begin{displaymath}
        \dl_{i} \equiv \mu^{2}[\dbar_{i}
                + (q^{-2} - q^{-1})X_{i}\dbar^{2}].
\end{displaymath}
$\dl_{i}$ then satisfy the relations
\begin{displaymath}
        [\dl \times \dl]_{k} = 0
\end{displaymath}
and
\begin{displaymath}
        \dl_{i}X_{j} = \imet{i}{j} +
q\Rinv{k}{l}{i}{j}X_{k}\dl_{l}
\end{displaymath}
The *-operation on $\mu$ and
$\dbar_{i}$ is defined as $\mu^{\ast} \equiv q^{-3}\invmu$
and $(\dbar_{i})^{\ast} \equiv -q^{3}\met{i}{j}\dl_{j}$.

Neither triplet of derivative operators has a subalgebra
isomorphic to \axr, but a linear combination of the two does.
This linear combination is then the triplet of \q-momentum
operators
\begin{displaymath}
        P_{i} = \frac{\dl_{i} +
                q^{-3}\dbar_{i}}{i(1 + q^{-3})}.
\end{displaymath}
Then $(P_{i})^{\ast} = \met{i}{j}P_{j}$,
$[P \times P]_{k} = 0$, and $P^{2}$ is a real scalar
that commutes with the $P_{i}$.

The $\X{i}$ and $P_{j}$ satisfy \q-deformed versions of the
Heisenberg relations:
\begin{displaymath}
        i(P_{a}X_{b} - q\Rinv{c}{d}{a}{b}X_{c}P_{d})
                = \invmu \left( \imet{a}{b}W +
                \frac{q - 1}{qK} \ieps{a}{b}{m}
                    L_{m} \right)
\end{displaymath}
and
\begin{displaymath}
        -i(X_{a}P_{b} - q\Rinv{c}{d}{a}{b}P_{c}X_{d})
                = q^{3}\mu \left( \imet{a}{b}W +
                \frac{q - 1}{qK} \ieps{a}{b}{m}
                L_{m} \right),
\end{displaymath}
where $W \equiv \mu[1 + q^{-2}(1 - q)X \cdot \dbar]$
is a real scalar,
$L_{i} \equiv \iq \mu [X \times \dbar]_{i}$,
and
\begin{equation}
          K \equiv q - 1 + \iq.
          \label{eq:Kq}
\end{equation}
The $L_{i}$ and $W$ generate a \q-deformed
angular momentum algebra.  Vectors $Z_{i}$ (such as the
$X_{i}$, $\dl_{j}$, and $P_{k}$)
satisfy the following relations with the $\L_{i}$ and $W$,
generalizing the role of the $L_{i}$ as generators
of rotations:
\begin{displaymath}
        L_{i}Z_{j} = - \eps{i}{c}{d}\ieps{d}{j}{e}Z_{c}L_{e}
                + \ieps{i}{j}{a}Z_{a}W,
\end{displaymath}

\begin{displaymath}
        L \cdot Z = Z \cdot L = 0,
\end{displaymath}

\begin{displaymath}
        WZ_{j} = KZ_{j}W - (K - 1)\eps{j}{r}{s}Z_{r}L_{s}.
\end{displaymath}
In addition $Z^{2}$ commutes with the $L_{i}$ and $W$.
The $L_{i}$ and $W$ also satisfy the following
relations:
\begin{displaymath}
        L^{2} = \frac{W^{2} - 1}{K - 1}
\end{displaymath}
and
\begin{displaymath}
        [L \times L]_{k} = L_{k}W = WL_{k}.
\end{displaymath}

\axr\ modulo powers of $X^{2}$ can be shown to be
a direct sum, indexed by nonnegative integers $l$, of
irreducible representations of the angular momentum
algebra.  The $l$th representation is then $(2l+1)$-dimensional,
and $W$ is a Casimir operator with eigenvalue
\begin{equation}
        w_{l} = \frac{q^{l+1} + q^{-l}}{q + 1}.
	\label{eq:wl}
\end{equation}
For $q=1$, the eigenvalue of $L^{2}$
in the $l$th representation becomes the familiar $l(l+1)$.

As a last preliminary result, the momentum operators
can be expressed in
terms of $W$, $\mu$, and $X^{2}$:
\begin{equation}
        X^{2}P_{i} = \frac{1}{iK(q - \iq)(q - 1)}
                [\iq X_{i}W\invmu - WX_{i}\invmu
                + q^{2}X_{i}W\mu - qWX_{i}\mu],
        \label{eq:X2Pi}
\end{equation}
and
\begin{equation}
        X^{2}P^{2} = \frac{1}{K^{2}(q - \iq)^{2}}
                [\frac{(q + 1)^{2}}{q}W^{2}
                - q\mu^{2} - 2 - \iq (\invmu)^{2}].
        \label{eq:X2P2}
\end{equation}
These identities
will be essential for calculations in the
representation we will introduce later.
We have included the numerical $K$ factors on the left hand side of these
identities, as we will define the hamiltonian scaled by a $K^2$ factor,
which
goes to one when $q \rightarrow 1$.

\section{Definition of \invR}

\invR\ is already a well-defined concept in the space
of undeformed, complex functions on {\bf $R^{3}$}.
Its essential properties are that it is a real, scalar
function and that
$X^{2}(\invR)^{2} = (\invR)^{2}X^{2} = 1$.  The simplest
generalization of these properties is then to
define \invR\ to be a real, scalar
corepresentation of \so .  \axre\ is then
the *-algebra generated by the $X_{i}$ and
\invR , where the \X{i} obey the same
relations as in \axr, \invR\ commutes with the \X{i},
and
\begin{equation}
        X^{2}(\invR)^{2} = 1.           \label{eq:iden}
\end{equation}

\dxre\ is the \so -covariant *-algebra of operators on
\axre, where we add \invR\ to the
generators of \dxre.  For this definition to be complete
we must have a set of relations involving \invR, $\mu$,
and the derivatives.  These must be consistent with
the relations between \invR\ and the \X{i}.  Clearly
since \invR\ has dimensions of an inverse length,
we should have
\begin{displaymath}
        \mu \invR = \qn{-1} \invR \mu .
\end{displaymath}
Equation \ref{eq:iden} implies
$\dl_{i}(\invR)$ may be.
\begin{displaymath}
        \dl_{i} X^{2}(\invR)^{2} = \dl_{i}(1) = 0
\end{displaymath}
It follows from the algebra of \dxr\ that
\[ \dl_{i} X^{2} = (q^{-1} + 1) X_{i}
          + q^{2}X^{2}\dl_{i}. \]
Thus
\[      (q^{-1} + 1) X_{i} \invRn{2}
        + q^{2}X^{2}\dl_{i}(\invRn{2}) = 0, \]
and therefore
\begin{equation}
        \dl_{i} \left( \invRn{2} \right)
        = -q^{-2} (q^{-1} + 1) \invRn{4} \X{i}
        \label{eq:d(invR)}
\end{equation}
Note that we are using the notation that if $A \in \dxre$,
and $f \in \axre$ then $A(f) \in \axre$ is the result of
evaluating the effect of $A$ on $f$, whereas $Af \in \dxre$
is the product of $A$ and $f$ as operators.

The simplest solution for $\dl_{i} \invR$ is
\begin{equation}
        \dl_{i} \invR = q^{-1} \invR \dl_{i}
        - q^{-2} \invRn{3} \X{i}.  \label{eq:dinvR}
\end{equation}
Repeated application of Equation (\ref{eq:dinvR}) indeed
gives (\ref{eq:d(invR)}).
Similarly one finds that
\begin{equation}
          \dbar_{i} \invR = q \invR \dbar_{i}
        - q^{2} \invRn{3} \X{i}.
	\label{eq:dbarinvR}
\end{equation}

Equation (\ref{eq:dinvR}) must be checked for consistency
with the algebra of \axre\ before we can conclude
that \dxre\ is a consistent operator algebra.  In
particular we must show that
$\dl_{i} (\invR f) = \dl_{i} (f \invR)$ and
$\dl_{i} (X^{2} \invRn{2}) = \dl_{i} (\invRn{2} X^{2}) = 0$.
In addition if \invR\ is truly a scalar, it should commute
with the $L_{i}$ and $W$.  The proofs that these conditions
are satisfied are given in Appendix \ref{proofs}.

Having defined \invR , we can now also define the
\q-deformed radius $R \equiv \invR X^{2}$.  This has
the following commutation relation with $\dl_{i}$:
\begin{eqnarray*}
        \dl_{i} R & = & \dl_{i} \invR X^{2} \\
        & = & \left[ \qn{-1} \invR \dl_{k}
- \qn{-2} \invRn{3} \X{k} \right] X^{2} \\
        & = & \qn{-1} \invR [(q^{-1} + 1) X_{i}
+ q^{2}X^{2}\dl_{i}] - \qn{-2} \invR \X{i} \\
        & = & \qn{-1} \invR \X{i} + qR \dl_{i}
\end{eqnarray*}
Induction over positive and negative n gives
\[      \dl_{i} R^{n} = \qn{-1} (n)_{q} R^{n-2} X_{i}
+ q^{n} R^{n} \dl_{i}   \]
where
\[      (n)_{q} \equiv \frac{\qn{n} - 1}{q - 1} \]
is the asymmetric \q-analog.  One can in fact develop
a theory using \dl\ or \dbar\ as momentum operators
and rewrite \q-deformed harmonic oscillator theories in
the language of $R$ and integral powers of $R$.  However
since $\dl^{2}$ and $\dbar^{2}$ are not self-adjoint, we
will concentrate on the action of the $P_{i}$ on
elements of \axre .

\section{Separation of Variables}

Using the formalism developed in the last section, we can now consider
the $q$-analog of the separation of variables problem for the kinetic term
of the hamiltonian. To this end we first introduce $q$-analogs of spherical
harmonics. These are defined, up to a---for us irrelevant---normalization
as elements \yq{l}{m} of \axre\, which obey the following two conditions:\\
(i) Left multiplication of \yq{l}{m} by $L^2$ or by $L_2$ yields \yq{l}{m}
multipied by a real eigenvalue, or in other words, \yq{l}{m} is an
``eigenfunction'' of both $L^2$ and $L_2$.\\
(ii) All \yq{l}{m} commute with $\mu$.\\

For illustration, we give the explicit expressions of \yq{l}{m} for
$l=0,1,2$
\begin{eqnarray*}
        \yq{0}{0} & = & 1 \\
        & & \\
        \yq{1}{-1} & = & \invR \X{1} \\
        \yq{1}{0} & = & \invR \X{2} \\
        \yq{1}{1} & = & \invR \X{3} \\
        & & \\
        \yq{2}{-2} & = & \invRn{2} \X{1}^{2} \\
        \yq{2}{-1} & = & \invRn{2} \X{1} \X{2} \\
        \yq{2}{0} & = & \invRn{2} [q\X{1}\X{3}
- (\sq + \isq)\X{2}\X{2} + \iq \X{3}\X{1}] \\
        \yq{2}{1} & = & \invRn{2} \X{2} \X{3} \\
        \yq{2}{2} & = & \invRn{2} \X{3}^{2}
\end{eqnarray*}

Now, by multiplying equations (\ref{eq:X2Pi}) and (\ref{eq:X2P2})
on the left by \invR\ and \invRn{2} respectively,
we obtain
\begin{equation}
        KP_{i} = \frac{1}{i(q - \iq)(q - 1)} \invR
        \left[ \iq X_{i}W\invmu - WX_{i}\invmu
+ q^{2}X_{i}W\mu - qWX_{i}\mu \right],
        \label{eq:QP}
\end{equation}
and
\begin{equation}
        K^{2}P^{2} = \frac{1}{(q - \iq)^{2}}
        \invRn{2} \left[\frac{(q + 1)^{2}}{q}W^{2}
        - q\mu^{2} - 2 - \iq (\invmu)^{2} \right].
        \label{eq:Q2P2}
\end{equation}

Since $\mu$ commutes with the \yq{l}{m} and $W$ commutes
with powers of $R$, if we expand functions in terms of
$R^{n}\yq{l}{m}$, Equation (\ref{eq:Q2P2}) is ready made
for calculating their momentum squared.
\begin{eqnarray*}
        K^{2}P^{2} \RY{n}{l}{m} & = & \frac{1}{(q - \iq)^{2}}
\invRn{2} \left[ \frac{(q + 1)^{2}}{q}w_{l}^{2}
- q^{2n+1} - 2 - q^{-2n-1} \right] \RY{n}{l}{m} \\
        & = & \frac{q^{2l+1} + q^{-2l-1} - q^{2n+1}
- q^{-2n-1}}{(q - \iq)^{2}} \RY{n-2}{l}{m}.
\end{eqnarray*}
In terms of the symmetric \q -analog of n of equation (\ref{eq:qan}),
we can write this in the simplified form
\begin{equation}
        K^{2} P^{2} \RY{n}{l}{m}
        = -\nq{n+l+1}\nq{n-l} \RY{n-2}{l}{m}.
        \label{eq:Q2P2R}
\end{equation}
This is a clear
generalization of the result from ordinary real calculus
that
\[
        \nabla [r^{n} Y_{lm}(\theta , \phi )] =
        (n + l + 1)(n - l) r^{n-2} Y_{lm}(\theta , \phi ),
\]
and it is the main result of this section.

Equation (\ref{eq:QP}) is not nearly as useful as its
counterpart because the action of the \X{i} on
\yq{l}{m} is nontrivial.  However for powers of $R$,
one can obtain the simple result that
\begin{equation}
        iKP_{i}(R^{n}) = \nq{n} R^{n-2} \X{i}
        = \nq{n} R^{n-1} \frac{\X{i}}{R}.
	\label{eq:KPRn}
\end{equation}

\section{The free particle}

Let us consider a system with Hamiltonian
\[      H = \QP        \]
with the convenient normalization factor $K^2$ determined by equation (3).
Then the Schr\"{o}dinger equation for this system
is the \q-deformed Helmholtz equation
\[
        \QP \psi = k^{2} \psi.
\]
The solutions to this are of the form
$j_{[q]l}(kR)\yq{l}{m}$ and
$n_{[q]l}(kR)\yq{l}{m}$ where $j_{[q]l}$ and $n_{[q]l}$
are respectively the \q-spherical Bessel and
Neumann functions.
\begin{equation}
        j_{[q]l}(x) = \sum_{n=0}^{\infty}
        \frac{(-1)^{n} \nq{2n+2l}!!}{\nq{2n}!! \nq{2n+2l+1}!} x^{2n+l}
	\label{eq:Bessel}
\end{equation}
\[
        n_{[q]l}(x) = - \sum_{n=0}^{l-1}
        \frac{\nq{2l-2n}!}{\nq{2n}!! \nq{2l-2n}!!} x^{2n-l-1}
\]
\begin{equation}
        + (-1)^{l+1} \sum_{n=l}^{\infty}
        \frac{(-1)^{n} \nq{2n-2l}!!}{\nq{2n}!! \nq{2n-2l}!}
        x^{2n-l-1}
	\label{eq:Neumann}
\end{equation}
where for nonnegative integers $n$
\begin{equation}
        \nq{n}! \equiv \left\{
	\begin{array}{ll}
	\prod_{k=1}^{n} \nq{k} & n > 0 \\
	1 & n = 0
	\end{array} \right.
	\label{eq:factorial}
\end{equation}
and
\begin{equation}
        \nq{2n}!! \equiv \left\{
	\begin{array}{ll}
	\prod_{k=1}^{n} \nq{2k} & n > 0 \\
      1 & n = 0
	\end{array} \right.
\end{equation}
That Equations (\ref{eq:Bessel}) and (\ref{eq:Neumann})
give rise to solutions is easily seen
by applying Equation (\ref{eq:Q2P2R}).  This eliminates the
zeroth element, and then reindexing gives the desired result
as would be the case for the $q=1$ differential equation.

We can also obtain a \q -deformed generalization
of the Rayleigh formulas, which provide an alternative
definition of these Bessel and Neumann functions.  This
will be a more convenient form for obtaining
\q-spherical Hankel functions,
\begin{equation}
        h_{[q]l}^{(1)}(x) \equiv j_{[q]l}(x) + i n_{[q]l}(x)
\end{equation}
and
\begin{equation}
        h_{[q]2}^{(1)}(x) \equiv j_{[q]l}(x) - i n_{[q]l}(x),
\end{equation}
corresponding to incoming and outgoing spherical waves.
These results are discussed in Appendix \ref{Rayleigh}.

It is interesting to note that, because the $P_{i}$ do not commute, it is
not
possible to have a plane wave with definite momentum as
in ordinary quantum mechanics.  The best we could
do is specify the component of the momentum in the direction
of propagation.  Since, moreover, $P_{2}$ for example does not
commute with $L_{2}$, the problem of expanding even
these quasiplane waves as a sum of \RY{n}{l}{m} terms
is nontrivial.

\section{The \q-Coulomb Problem}

In ordinary quantum mechanics, the Coulomb
Hamiltonian is
\[
        H = \frac{p^{2}}{2m} - \frac{\alpha}{r};
\]
or if we rescale this by $2m$ and incorporate the
mass into $\alpha$,
\[
        H = p^{2} - \frac{2\alpha}{r}.
\]
There are several possible ways to \q-deform this.
We are interested in a self-adjoint Hamiltonian which
preserves the properties of the ordinary Hamiltonian that
make it amenable to finding eigenfunctions.  That
is to say we require the existence of a \q-deformed Lenz vector
which commutes with the \q-deformed Hamiltonian so that
there continue to be degeneracies between solutions
with different angular momentum quantum numbers.

Following \cite{Weich}, we define our \q-deformed
Coulomb Hamiltonian to be:
\[
        H = \QP - \alpha q(\invR \mu + \mu^{\ast} \invR)
\]
which is clearly self-adjoint (for convenience the normalization of the
kinetic term has again been chosen with a prefactor $K^2$
determined by equation (3)). Noting that
$\mu^{\ast} = \qn{-3} \invmu$, this can be written
in the simpler form
\begin{equation}
        H = \QP - \alpha \invR (q \mu + \qn{-1} \invmu).
        \label{eq:ham}
\end{equation}
This Hamiltonian also commutes with the Lenz vector
\begin{eqnarray*}
        A_{k} & \equiv &
        \frac{[W, iKP_{k}]}{K - 1} + \frac{\alpha \X{k}}{R}
\\
        & = & iK(P_{k}W - (P \times L)_{k})
        + \frac{\alpha \X{k}}{R}.
\end{eqnarray*}
The Lenz vector along with the angular momentum operators
then generate the algebra given in \cite{Weich}.

If we write the eigenvalues $E$ of the Hamiltonian (\ref{eq:ham}) in
the form
\begin{equation}
E = - \left( \frac{\alpha}{\nq{\gamma}} \right)^{2},
\label{eq:eig}
\end{equation}
then the corresponding ``eigenfunctions'' are (see Appendix
\ref{Coulomb} for their derivation)
\begin{equation}
        \psi_{\gamma lm}
        = \sum_{p=0}^{\infty} A_{p}(\gamma) R^{l}
          \left( \frac{\alpha \qn{\ga}R}{\nq{\gamma}} \right)^{p}
        \expq \left( \frac{- \qn{l+1+p-\gamma}
        \alpha R}{\nq{\gamma}} \right) \yq{l}{m}
        \label{eq:psi}
\end{equation}
with
\begin{equation}
      A_{p}(\gamma) = \qn{pl+\half p(p+1)}
        \frac{(1-\qn{2})^{p}}{\poch{\qn{2}}{\qn{2}}{p}}
        \frac{\poch{\qn{2(l+1-\ga)}}{\qn{2}}{p}}
        {\poch{\qn{2(2l+2)}}{\qn{2}}{p}}
      \frac{\poch{\qn{4(l+1)}}{\qn{4}}{p}}
        {\poch{\qn{2(l+1)}}{\qn{2}}{p}}.
      \label{eq:Ap}
\end{equation}
Here
\begin{equation}
        (a;u)_{p} \equiv \left\{
	\begin{array}{ll}
	\prod_{m=0}^{p-1} (1 - a u^{m}) & p = 1, 2, 3 . . . \\
	1 & p = 0
	\end{array} \right.,
\end{equation}
is the \q-deformed Pochhammer symbol \cite{GR}, with
$u = \qn{2}$ and $\qn{4}$ in Eq. (\ref{eq:Ap}), and
\begin{equation}
        \expq (x)=\sum_{n=0}^{\infty} \frac{x^n}{[n]_q!}
\end{equation}
is the \q-deformed exponential, where we have used the notation
of Eq. (\ref{eq:factorial}).
$q^{\gamma}$ is obtained in terms of
the energy E from the quadratic equation (\ref{eq:eig}).
This has two solutions
\begin{equation}
q^{-\gamma_{\pm}} = \frac{\eta \pm \sqrt{\eta^{2} -E}}{\sqrt{-E}},
\label{eq:sol}
\end{equation}
with $\eta$ given by
\begin{equation}
        \eta = \frac{(\qn{-1} - q)\alpha}{2}.
	\label{eq:eta}
\end{equation}
(We assume $q < 1$ to insure convergence; had we instead set
$q > 1$, we would have to everywhere change $q \rightarrow \qn{-1}$)

Note that for a given energy, we actually only have one solution since
if we write the solutions as power series of $R$, the difference
equation admits only one solution that behaves as $R^{l}$ for small $R$.
Whether we write it in terms of $\ga_{+}$ or $\ga_{-}$ simply gives us
two expressions for the same result.

In ordinary quantum mechanics, a decaying exponential
multiplied by a polynomial in $r$ is normalizable and gives
rise to a bound state.  Something very similar happens in the
$q \neq 1$ limit.  Consider the wave function (\ref{eq:psi}), which
is a sum of terms each of which is a nonnegative power of R
multiplying a \q-exponential of the form $\expq (-c_{q}(p)R)$.
As can be seen from (\ref{eq:psi}), the $c_{q}(p)$ become
independent of $p$ as $q \rightarrow 1$, thus reproducing the
usual result.  Then we define a bound state in the \q-deformed
case to be a wave function of the type (\ref{eq:psi}) for
which the sum over $p$ truncates at the finite value $p = n-l-1$.
On account of the factor $\poch{\qn{2(l+1-\ga)}}{\qn{2}}{p}$ in
the numerator of (\ref{eq:Ap}), such a truncation occurs if
$\ga$ equals a positive integer $n > l$.  From Eq. (\ref{eq:eig}),
we then obtain the \q-Balmer formula, and we find the corresponding
wave functions given by (\ref{eq:psi}) and (\ref{eq:Ap}).  The \q-Balmer
formula already
appears in Ref. \cite{Weich}, where, as noted in the introduction,
the operator \invR  is treated differently.  This difference is
reflected in our wave functions.

In principle these same wave functions (\ref{eq:psi}) should
also cover the continuum part of the spectrum, and one should
be able to extract an $S$-matrix from them.  How to do this
in a rigorous fashion remains to be seen.  Here we consider
the candidate $S$-matrix suggested by equation (24)
\begin{equation}
        S_{l}^{(q)}(E) = (1 - \qn{2})^{(\ga_{-} - \ga_{+})}
        \frac{\Gamma_{\qn{2}} (l + 1 - \ga_{+})}
        {\Gamma_{\qn{2}} (l + 1 - \ga_{-})}=
\prod_{n=0}^{\infty} \frac{1-q^{2(l+1- \ga{-}+n)}}{1- q^{2(l + 1- \ga{+}
+n)}},
        \label{eq:S}
\end{equation}
where the \q-gamma function is defined as in \cite{GR}:
\begin{equation}
\Gamma_{q^2}(x):=
\frac{(q^{2};q^{2})_\infty}{(q^{2x};q^{2})_\infty} (1-q^{2})^{1-x}.
\end{equation}
The $S$-matrix (\ref{eq:S}) appears to have all the right
features:

A. For $q \rightarrow 1$ this $S$-matrix reproduces the familiar Coulomb
$S$-matrix. In this limit the prefactor goes to one , as can be seen from
Eq. (\ref{eq:sol}), and the $q$-gamma functions become precisely the
ordinary gamma functions which appear in the ordinary Coulomb $S$-matrix.

B. For integer $\ga_{+} \geq l+1$, $S_{l}^{(q)}(E)$ has a pole corresponding to
a $q$-Balmer state. Both the the location and the residue of this pole
differ
from those of its ordinary ($q=1$) Balmer limit.

C. As can be seen from Eqs. (\ref{eq:S}) and (\ref{eq:sol}) the
branchpoint of $S^{(q)}_{l}(E)$, the scattering threshold, is now located
$E=\eta^2$,
with $\eta$ given in Eq. (\ref{eq:eta}) and not at $E=0$ as
in the ordinary case.  This is the most dramatic departure from the
ordinary case: the scattering
region starts at $E=\eta^2$. For $q=1$ this reduces to the expected
threshold $E=0$.

\appendix

\section{Properties of \invR}
\label{proofs}

We have already satisfied $\dl_{i} (X^{2} \invRn{2}) = 0$.
This definition of $\dbar_{i} \invR$ also satisfies the
other half of the second constraint.
\begin{displaymath}
        \dl_{i} \invRn{2} = q^{-2} \invRn{2} \dl_{i}
        - q^{-2} (q^{-1} + 1) \invRn{4} \X{i}
\end{displaymath}

\begin{displaymath}
        \dl_{i} \left( \invRn{2} X^{2} \right)
         = \left[ q^{-2} \invRn{2} \dl_{i} %
+ \qn{2} X^{2} \dl_{i} \right] (1) %
- \qn{-2} (\qn{-1} + 1) \invRn{2} \X{i}
\end{displaymath}
\begin{displaymath}
 = 0
\end{displaymath}

Trivially $\dl_{i} 1 \invR = \dl_{i} \invR 1 = \dl_{i} \invR$.
Suppose that $\dl_{i} f \invR = \dl_{i} \invR f$ where
$f \in \axre$.
\begin{eqnarray*}
        \dl_{i} \X{j} f \invR & = & %
[\imet{i}{j} + q\Rinv{k}{l}{i}{j} \X{k} \dl_{l}]f \invR \\
        & = & \imet{i}{j} f \invR %
+ q \Rinv{k}{l}{i}{j} \X{k} \dl_{l} \invR f \\
        & = & \imet{i}{j} f \invR %
+ q \Rinv{k}{l}{i}{j} \X{k} %
\left( \qn{-1} \invR \dl_{l} - \qn{-2} \invRn{3} \X{l} \right) f \\
        & = & \imet{i}{j} f \invR \\
        & & -\qn{-2} \kron{k}{i} \kron{l}{j} %
                \invRn{3} \X{k} \X{l} f %
                + \qn{-1}\eps{a}{k}{l}\eps{i}{j}{a}\X{k}\X{l} %
                f \invRn{3} \\
        & & -\qn{-1} (q - 1) \met{k}{l} \imet{i}{j}
\invRn{3}\X{k}\X{l}f \\
        & = & \qn{-1}\imet{i}{j} f \invR
+ \invR \Rinv{k}{l}{i}{j} \X{k} \dl_{l} f
\end{eqnarray*}
where we used the fact that $\eps{a}{k}{l}\X{k}\X{l} = 0$.
At the same time
\begin{eqnarray*}
        \dl_{i} \invR \X{j} f & = & \left(q^{-1} \invR \dl_{i}
- q^{-2} \invRn{3} \X{i} \right) \X{j} f \\
        & = & \qn{-1} \invR [\imet{i}{j} +
q \Rinv{k}{l}{i}{j} \X{k} \dl_{l}] f
- \qn{-2} \invRn{3} \X{i} \X{j} f \\
        & = & \dl_{i} \X{j} f \invR
\end{eqnarray*}
We must also show that
$\dl_{i} \invR f \invR = \dl_{i} f \invR \invR$ in
order to complete this inductive proof.  However this
is trivial having assumed that
$\dl_{i} f \invR = \dl_{i} \invR f$.
Thus \axre\ and \dxre\ are consistently defined.

If \invR\ is truly a scalar, it should commute with
the $L_{i}$ and $W$.  This is indeed true:
\begin{eqnarray*}
        W & = & \mu [1 + \qn{-2}(1 - q) X \cdot \dbar] \\
        X \cdot \dbar & = & \met{i}{j} \X{i} \dbar{j} \invR \\
        & = & \met{i}{j} \X{i} \left[ q \invR \dbar_{j}
- \qn{2} \invRn{3} \X{j} \right] \\
        & = & q \invR X \cdot \dbar - \qn{2} \invR \\
        W \invR & = & \mu \left[ \invR
+ (\qn{-2} - \qn{-1}) \left( q \invR X \cdot \dbar
- \qn{2} \invR \right) \right] \\
& = & \mu \left[ q \invR
+ q \invR (\qn{-2} - \qn{-1}) X \cdot \dbar \right] \\
          & = & \invR \mu [1 + (\qn{-2} - \qn{-1}) X \cdot \dbar] \\
        & = & \invR W
\end{eqnarray*}

\begin{displaymath}
        L_{i} = \mu \qn{-1} [X \times \dbar]_{i}
\end{displaymath}
\begin{eqnarray*}
        [ X \times  \dbar ]_{i} & = & \eps{i}{j}{k} \X{j} \dbar_{k} \\
& = & \eps{i}{j}{k} \X{j} \left[ q \invR \dbar_{k}
                - \qn{2} \invRn{3} \X{k} \right] \\
        & = & q \invR \eps{i}{j}{k} \X{j} \dbar_{k} \\
        & = & q \invR [X \times \dbar]_{i} \\
        L_{i} \invR & = &
\mu \qn{-1} [X \times \dbar]_{i} \invR \\
        & = & \mu q \invR \qn{-1} [X \times \dbar]_{i} \\
        & = & \invR \mu \qn{-1} [X \times \dbar]_{i} \\
        & = & \invR L_{i}
\end{eqnarray*}

\section{\q-deformed Rayleigh formulas}
\label{Rayleigh}

Let $D$ be the symmetric \q-derivative:
\begin{equation}
        Df(x) \equiv \frac{f(qx) - f(\qn{-1}x)}
        {(q - \qn{-1})x}
\end{equation}
Acting on monomials,
\begin{equation}
        Dx^{n} = \nq{n} x^{n-1}
\end{equation}

Let us define the following \q-generalizations of some
common functions by replacing factorials with \q-deformed
factorials in their Taylor expansions:
\begin{eqnarray}
        \expq (x) & \equiv & \sum_{n=0}^{\infty}
        \frac{x^{n}}{\nq{n}!}
\\
        \cosq (x) & \equiv & \sum_{n=0}^{\infty}
        \frac{(-1)^{n} x^{2n}}{\nq{2n}!}
\\
        \sinq (x) & \equiv & \sum_{n=0}^{\infty}
        \frac{(-1)^{n} x^{2n+1}}{\nq{2n+1}!}.
\end{eqnarray}
Then $\expq (ix) = \cosq (x) + i\, \sinq(x)$

The \q-deformed Rayleigh formulas,
\begin{eqnarray}
        j_{[q]l}(x) & = & (-x)^{l} \left( \frac{1}{x}
        D \right)^{l} \left( \frac{\sinq (x)}{x} \right)
\\
        n_{[q]l}(x) & = & (-x)^{l} \left( \frac{1}{x}
        D \right)^{l} \left( \frac{-\cosq (x)}{x} \right),
\end{eqnarray}
can then be proved by induction after noting that
\begin{equation}
      j_{[q]0}(x) = \frac{\sinq (x)}{x}
\end{equation}
and
\begin{equation}
      n_{[q]0}{x} = - \frac{\cosq (x)}{x}.
\end{equation}
The \q-deformed spherical Hankel functions are
\begin{equation}
        h_{[q]l}^{(1)}(x) \equiv j_{[q]l}(x) + i n_{[q]l}(x)
\end{equation}
and $h_{[q]l}^{(2)}(x)$ is just the complex conjugate of
$ h_{[q]l}^{(1)}(x)$.  Since $(-x)^{l}(\frac{1}{x} D)^{l}$ is a linear
operator, it follows that $h_{[q]l}^{(1)}$ should
satisfy
\[
        h_{[q]l}^{(1)}(x) = (-x)^{l} \left( \frac{1}{x}
        D \right)^{l} \left( \frac{-i \expq (ix)}{x} \right).
\]
One can show by induction that
\begin{equation}
        h_{[q]l}^{(1)}(x) = \sum_{n=0}^{l}
        q^{\frac{1}{2} [l(l+1)-n(n+1)]}
        \frac{i^{n-l-1} \nq{l+n}!}{\nq{2n}!! \nq{l-n}!}
        \frac{\expq (iq^{n}x)}{x^{n+1}}
\end{equation}
satisfies this Rayleigh formula for all $l$.  Thus these
functions must equal the Hankel functions.

The powers of $q$ that appear both inside and outside the
q-exponential arise because
\begin{eqnarray*}
        D(x^{k} \expq (\alpha x))
        & = & \nq{k} x^{k-1} \expq (\alpha qx)
        + \alpha q^{-k} x^{k} \expq (\alpha x) \\
        & = & \nq{-k} x^{k-1} \expq (\alpha \iq x)
        + \alpha q^{k} x^{k} \expq (\alpha x).
\end{eqnarray*}
This results from the \q-deformed arithmetic in which
\begin{equation}
        \nq{n+k} = q^{n}\nq{k} + q^{-k}\nq{n}
        = q^{-n}\nq{k} + q^{k}\nq{n}
\end{equation}
It appears to be a common trend of solutions
to self-adjoint \q-deformed Hamiltonians that they
can be expressed as series of the form
$\sum_{n} A_{n} \expq (q^{n}x)$.

\section{\q-Coulomb Hamiltonian Eigenvalue Problem}
\label{Coulomb}

Let $\beta = \sqrt{-E}$ and
$\nq{\gamma} = \frac{\alpha}{\beta}$.
In order to obtain
a difference equation for the $A_{p}$, we need to
express $(H + \beta^{2}) \psi$ as a series of
the form $\sum_{p} B_{p} \expq (q^{p}x)$ where
$B_{p}$ is a function of the $A_{p}$.

If we simply apply $(H + \beta^{2})$ to $\psi$ using
Equation \ref{eq:Q2P2R}, we get
\begin{displaymath}
        (H + \beta^{2}) \Rexp{p}{s} =
\end{displaymath}
\[
        - \suminf{n} \frac{\nq{n+p+2l+1} \nq{n+p}}{\nq{n}!}
        (-q^{s}\beta )^{n}\RY{n+p+l-2}{l}{m}
\]
\[
         -  \alpha \suminf{n}
        \frac{(\qn{n+l+p-1} + \qn{-n-l-p-1})}{\nq{n}!}
        (-q^{s}\beta)^{n}\RY{n+p+l-1}{l}{m}
\]
\[
        + \beta^{2} \Rexp{p}{s}.
\]

If we decompose $\nq{n+p+2l+1} \nq{n+p}$, we get
terms proportional to \\
$\nq{n} \nq{n-1}$, $\qn{-n} \nq{n}$,
and $\qn{-2n}$.  By resumming these terms, we can
rewrite them as powers of R times exponentials.
We wish
to rewrite the entire equation in terms of functions
of the form \Rexp{p-m}{s+m} where $m$ is an integer and
$s$ is a function of $p$.  Then
we can obtain a difference equation for the coefficients
of these functions.
In the following, we assume that $s = -l-1-p+\gamma$.
The only terms which need to be rewritten are then
\begin{displaymath}
        (\QP - \alpha \invR \qn{-1} \invmu)
        \Rexp{p}{-l-1-p+\gamma} =
\end{displaymath}
\begin{displaymath}
        - \suminf{n} \frac{\nq{n+2l+p+1} \nq{n+p}}{\nq{n}!}
        (\qn{l+1+p-\gamma} \beta)^{n} \RY{n+l+p-2}{l}{m}
\end{displaymath}
\[
        - \nq{\gamma} \beta \suminf{n}
        \frac{\qn{n+l+p+1}}{\nq{n}!}
        (-\qn{l+1+p-\gamma} \beta)^{n} \RY{n+l+p-1}{l}{m}
\]
\[
        = - \suminf{n} \left( \nq{n+2l+p+1} \nq{n+p}
        - \qn{-1+n+\gamma} \nq{n} \nq{\gamma} \right)
\]
\[
        \times \frac{(-\qn{l+1+p-\gamma} \beta)^{n}}{\nq{n}!}
        \RY{n+l+p-2}{l}{m}
\]
\[
        = - \suminf{n} \left\{ (\qn{2p+2l+2} + 1
        - \qn{2\gamma}) \nq{n} \nq{n-1}
        + \qn{-2n} \nq{p+2l+1} \nq{p} \right.
\]
\[
        + \left. \qn{-n} \nq{n} \left( -\qn{1+\ga} \nq{\ga}
        \qn{p+2l+1} \nq{p} + \qn{1+p} \nq{2l+2+p} \right)
        + \qn{-2n} \nq{p+2l+1} \nq{p} \right\}
\]
\[
         \times  \frac{(-\qn{l+1+p-\gamma} \beta)^{n}}{\nq{n}!}
        \RY{n+l+p-2}{l}{m}
\]
\[
         =  \beta^{2} \left( \qn{2l+2+2p} - \qn{2l+2+2p-2\ga}
        - \qn{4l+4p+4-2\ga} \right) \Rexp{p}{-l-1-p+\ga}
\]
\[
         -  \beta \left( \qn{l+p+1} \nq{\ga}
        - \qn{2p+3l+1-\ga} \nq{p}
        - \qn{l+2p+1-\ga} \nq{2l+p+2} \right)
\]
\[
        \times \Rexp{p-1}{-l-p+\ga}
\]
\[
         -  \nq{p+2l+1} \nq{p} \Rexp{p-2}{-l-p+1+\ga}
\]
Thus
\begin{displaymath}
        (H + \beta^{2}) \Rexp{p}{-l-p-1+\ga}
\end{displaymath}
\begin{displaymath}
        = \qn{2l+2p+2-\ga} [2]_{q^{-l-p-1}}
        \nq{l+p+1-\ga} (\qn{-1} - q)
\end{displaymath}
\[
        \times \beta^{2} \Rexp{p}{-l-1-p+\ga}
\]
\[
        + \left( \qn{2p+3l+1-\gamma} \nq{p}
        + \qn{l+2p+1-\ga} \nq{2l+p+2}
        - [2]_{q^{l+p+1}} \nq{\ga} \right)
\]
\[
        \times \beta \Rexp{p-1}{-l-p+\ga}
\]
\[
        - \nq{p+2l+1} \nq{p} \Rexp{p-2}{-l-p+1+\ga}
\]
If we sum over $p$ from $0$ to $\infty$, we can obtain the
desired difference equation.  We normalize the wave
functions so that $A_{0} = 1$.  Then
\[
        A_{1} = \qn{l+1} \beta \frac{[2]_{\qn{l+1}}
        \nq{l+1-\ga}}{\nq{2l+2}}
\]
and for $p > 0$,
\begin{displaymath}
        \left\{ \nq{p+2l+3} \nq{p+2} \right\} A_{p+2}
\end{displaymath}
\[
        = \qn{2l+2p+2-\ga} [2]_{\qn{l+p+1}}
        \nq{l+p+1-\ga} (\qn{-1} - q) \beta^{2} A_{p}
\]
\[
        + \left( \qn{2p+3l+3-\ga} \nq{p+1}
        + \qn{l+2p+3-\ga} \nq{2l+p+3} - [2]_{\qn{-l-p-2}}
        \nq{\ga} \right) \beta A_{p+1}
\]
Equation \ref{eq:Ap} is then a solution to this
difference equation.  Thus the theorem is proved.

\end{document}